# Wafer-scale graphene field-effect transistor biosensor arrays with monolithic CMOS readout

*Miika Soikkeli[1]\*, Anton Murros[1], Arto Rantala[1], Oihana Txoperena[2], Olli-Pekka Kilpi[1], Markku Kainlauri[1], Kuura Sovanto[1], Arantxa Maestre[2], Alba Centeno[2], Kari Tukkiniemi[1], David Gomes Martins[1], Amaia Zurutuza[2], Sanna Arpiainen[1] and Mika Prunnila[1]*

[1]VTT Technical Research Centre of Finland Ltd, P.O. Box 1000, FI-02044 VTT, Espoo, Finland

[2]Graphenea Semiconductor SLU, Paseo Mikeletegi 83, 20009-San Sebastian, Spain

**ABSTRACT** The reliability of analysis is becoming increasingly important as point-of-care diagnostics are transitioning from single analyte detection towards multiplexed multianalyte detection. Multianalyte detection benefits greatly from complementary metal-oxide semiconductor (CMOS) integrated sensing solutions, offering miniaturized multiplexed sensing arrays with integrated readout electronics and extremely large sensor counts. The development of CMOS back end of line integration compatible graphene field-effect transistor (GFET) based biosensing has been rapid during the last few years, both in terms of the fabrication scale-up and functionalization towards biorecognition from real sample matrices. The next steps in industrialization relate to improving reliability and require increased statistics. Regarding functionalization towards truly quantitative sensors and on-chip bioassays with improved statistics require sensor arrays with reduced variability in functionalization. Such multiplexed bioassays, whether based on graphene or on other sensitive nanomaterials, are among the most promising technologies for label-free electrical biosensing. As an important step towards that, we report wafer-scale fabrication of CMOS integrated GFET arrays with high yield and uniformity, designed especially for biosensing applications. We demonstrate the operation of the sensing platform array with 512 GFETs in simultaneous detection for sodium chloride concentration series. This platform offers a truly statistical approach on GFET based biosensing and further to quantitative and multi-analyte sensing. The reported techniques can also be applied to other fields relying on functionalized GFETs, such as gas or chemical sensing or infrared imaging.

# 1. Introduction

Point-of-care (PoC) diagnostics is transitioning from single analyte detection towards multiplexed multianalyte detection which increases the importance of reliable statistical analysis[1]. When targeting quantitative or multianalyte sensors, simultaneous screening of a sample for several different analytes is required, together with adequate statistics to be able to exclude individual false signal sources. Traditionally multiplexed testing is done on enzyme-linked immunosorbent assays (ELISA)[2], where every analyte and receptor pair is measured individually. More advanced multiplexed multianalyte diagnostics solutions can be achieved on a single chip by using sensor arrays with integrated complementary metal-oxide semiconductor (CMOS) readout. The benefits of CMOS integration include low cost, dense array formation, lower power consumption, label-free detection mechanism, readout integration and smaller device dimensions[3,4] which all are very beneficial especially for PoC applications.

One of the most promising label-free biosensing technology, compatible with CMOS integration, is sensitive and selective graphene field-effect transistor (GFET) sensors based on chemical vapour deposited (CVD) graphene. The use of GFET has been demonstrated for peptides and antibodies[5], attomolar level DNA hybridization[6], zika-virus detection[7] and recently also for COVID-19 causative virus monitoring[8,9]. With CMOS integrated GFET arrays, it is possible to overcome the common problems in field effect biosensing related to the limited detection ranges in real-life sample matrices by utilizing PEG-polymer aided functionalization schemes[9,10]. This is all very promising considering the prospects of GFETs in label-free electrical sensing for early diagnostics and PoC applications requiring sensitive and specific biorecognition from biological sample matrices.

Other promising, compatible with CMOS back end of line (BEOL) integration, FET based technologies for biosensing are based on carbon nanotube (CNT)[11], organic semiconductor[12] and semiconductor silicon nanowire (SiNW)FET sensors[13]. All these technologies still have some limitations that need to be addressed to get the full potential out. Organic FETs have lower charge sensitivity and dynamic range when compared to GFETs and SiNWFETs due to larger channel dimensions.[14] SiNWFETs suffer from high device-to-device variations and low carrier mobility.[15] CNTFETs have low current output, small active areas and heterogenous semiconductor and metal mixtures.[16] These aspects can lead to a large variation between devices and low manufacturing yields for CNTFETs.[16] GFETs offer at least similar sensitivity and easier fabrication when compared to the competing technologies. This makes GFETs very interesting and cost-effective approach for highly sensitive label-free biosensors.

Previously, scalable CMOS integrated biosensor solutions have been reported for example with carbon nanotubes[16], ion selective FETs (ISFETs)[17], extended gate field effect transistors (EGFETs)[18] and film bulk acoustic resonators (FBARs)[19]. ISFETs and EGFETs are easy to manufacture on top of the CMOS with standard processes.[17,18] ISFETs also suffer typically from device stability and drift issues.[20,21] EGFETs were originally proposed as a solution to solve the drift and stability issues of ISFETs and the sensitivity of the devices can be improved due to the possibility for larger sensing areas.[22,23] CNT based sensors and FBARs are harder to fabricate on CMOS readout and easily suffer from low yields in fabrication.[16,19] With CNT based sensors, the as-grown angle of CNTs is difficult to control without deliberate alignment, and deposited films show high variability in contact resistance.[24] FBARs suffer from low yields due to residual stresses in the suspended membrane and piezoelectric layer and lengthy etching processes.[25] The possibility to integrate CVD graphene with CMOS readout on chip scale has been demonstrated earlier for broadband image detectors[26] and gas sensing[27]. The possibilities of graphene integration into CMOS back end of line (BEOL) have been discussed by Neumaier et al.[28] However, full BEOL integration of the graphene-based biosensor arrays have not been realized due to the lack of wafer-scale CMOS readout integration.

In this work, we demonstrate for the first-time wafer-scale CMOS integration of graphene FETs for biosensing. Functionality of the integrated GFETs for biosensing is demonstrated using simultaneous measurement of 512 GFETs for a series of sodium chloride concentration measurement. In our approach, we integrate GFETs on a CMOS multiplexer platform that enables simultaneous measurement of hundreds of GFETs. The work done here further enables a statistical approach for multi-analyte biosensing since the reliable detection of a certain analyte requires several devices and statistics for the analysis.

## 2. Materials and methods

**CMOS read-out design**

The CMOS wafers used for GFET integration were manufactured using a standard commercial technology provided by XFAB. The 200 mm CMOS wafers utilize a 0.35 µm analog CMOS process node with up to four metallization layers. The CMOS technology includes a wide range of active devices and high-performance analog devices.

The CMOS read-out platform was designed to support several types of post-processed sensors. The architecture of the CMOS system and the local readout for a resistive GFET is illustrated in Figure 1. The Application Specific Integrated Circuit (ASIC) is comprised of a macro array consisting of 64 sensor blocks. Each sensor block is further comprised of a sub-array with 64 individual sensors with four different read-out schemes. Each macro array thus contains readouts for up to 4096 individual devices. Here we focus only on the devices with analog front end (AFE) designed for the resistive read-out of the GFETs.

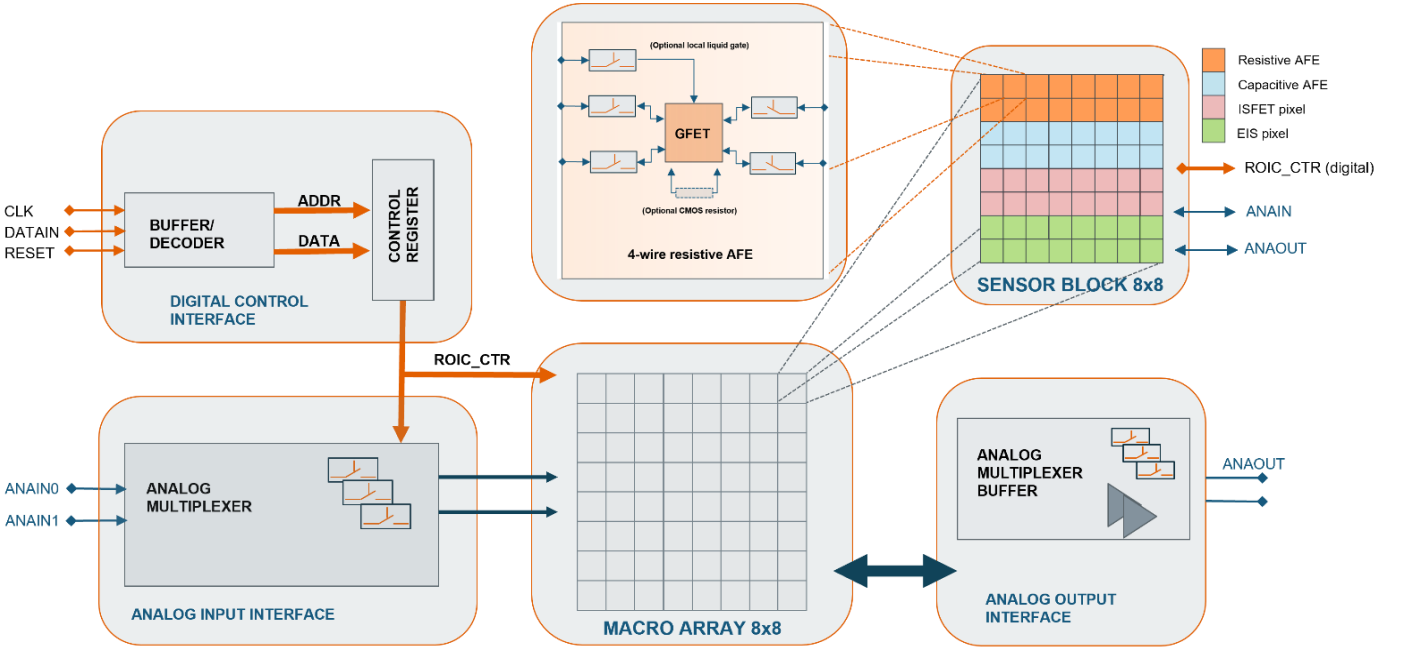

**Figure 1. Block diagram of the CMOS system with a single large read-out matrix having common global level control and output signal processing units. Different read-out electronics have been laid out around matrix area to obtain uniform coverage for each sensor type. The different sensor read-out circuits are grouped into an 8x8 sub-matrices. In the whole ASIC there is a macro array consisting of 8x8 sub-arrays. Each sensor can be individually selected. Inside each sub-array there are 2 rows of GFET sensors, capacitive graphene sensors, n- and p-type ISFET sensors and EIS sensors. The global digital control can be used for the resistive AFE of the GFETs to select a plurality of GFETs by enabling the pixel level local CMOS switches. There is an option to select either 2-wire or 4-wire connection to enable a kelvin type resistance measurement. A local gate electrode is also possible to connect to provide a bias for back or top gate electrode depending on process selection.**

The global digital control is used to select the GFETs by enabling pixel level local CMOS-switches. For each GFET there is an option to select either 2-wire or 4-wire connection to enable a kelvin type resistance measurement. The global control for the resistive GFETs operates as follows. The digital control SPI-interface is utilized to fetch a code string to address GFET(s) to be analyzed. The address decoder selects the corresponding GFET(s) and enables analog switches and the output multiplexer. The multiplexer output is in the case of the GFET read-out fed directly to ASIC pads. In this ASIC design the GFET resistance is measured using external laboratory equipment. CMOS readout also enables the integration of the resistance meter functionality on chip level which will be evaluated in the future designs to increase the functionality of the system.

**CMOS post-processing of the graphene field effect transistors**

The post-processed sensors were designed to utilize half of the available readouts to allow space for local liquid gate electrodes, and in addition to an on-chip global liquid gate electrode. These liquid gate electrodes can gate the graphene channels in biosensing measurements. Thus, the selected options result in 512 individual GFETs being fabricated on each chip. A microscope image of the fabricated GFET sensors is shown in Figure 2a.

A schematic illustration of the post-processed GFET on the CMOS readout is presented in Figure 2b. Post-processing was started by filling the CMOS vias by Pt deposited by sputtering and patterned using a lift-off process. All post-process patterning steps were done by standard UV-lithography. In addition of the via contacts the Pt metal layer also defined the on-chip Pt liquid gate that is used to gate the graphene channel. In the next step, CVD graphene was transferred on the CMOS wafers by using a wet transfer method. After the graphene transfer, the wafers were annealed 16 hours at 300 °C in vacuum. 50-nm-thick AuPd alloy contact metal was deposited by evaporation and lift-off. Next the graphene surface was protected with a 50-nm-thick $Al_2O_3$ grown by atomic layer deposition (ALD). ALD growth on graphene was seeded using an evaporated 1-nm-thick Al film oxidized in the evaporation chamber.

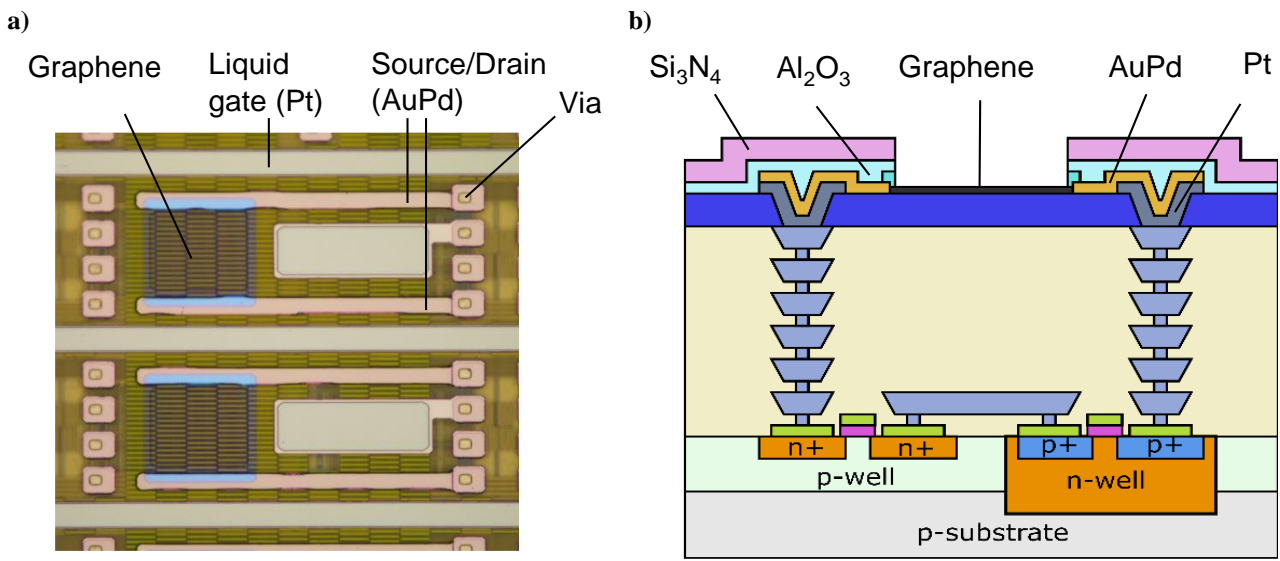

**Figure 2**. a) Graphene FETs on top of the CMOS readout circuit connected to the vias on the right side. The vias on the left side are not connected to leave more space for local and global liquid gate options. b) **A schematic illustration of the post-processed GFET on CMOS readout. Pt has been used to fill the vias and to form a liquid gate mesh on the surface of the chips. Contacts between graphene and the vias have been formed with AuPd metals. Graphene has been patterned by using a $Al_2O_3$ protective layer. The device has been passivated with $Al_2O_3$ and $Si_3N_4$ layers leaving only the graphene channel and liquid gate electrode surfaces open for the liquid measurements.**

Graphene patterning was done by wet etching the protective Al$_2$O$_3$ in H$_3$PO$_4$ followed by etching of the graphene in O$_2$ plasma. After the graphene patterning, the devices were passivated by a 50-nm-thick ALD Al$_2$O$_3$ and 100-nm-thick plasma-enhanced chemical vapor deposition (PECVD) Si$_3$N$_4$. Finally, the passivation of the devices was opened from the top of the graphene channels by dry etching the Si$_3$N$_4$ and wet etching the Al$_2$O$_3$. The wafers were then diced into single CMOS microsystem chips.

**Graphene FET measurements**

For the characterization, five sensor chips were selected across the diced wafer. The chips were wire bonded onto chip carriers for electrical measurements. A detector chip wire bonded on a chip carrier is shown in Figure 3a. For one of the chips, the bonded wires were protected with polydimethylsiloxane (PDMS) coating to enable a measurement in liquid. Electrical measurement configuration for the GFETs is shown in Figure 3b. A parameter analyzer is used together with the CMOS multiplexer on the chips for the measurement and biasing of the devices. Liquid gate voltage ($V_G$) is controlled by using the on-chip Pt liquid gate. The current (I) in the graphene channel is measured with 0.1 V source-drain bias ($V_{SD}$) over the channel.

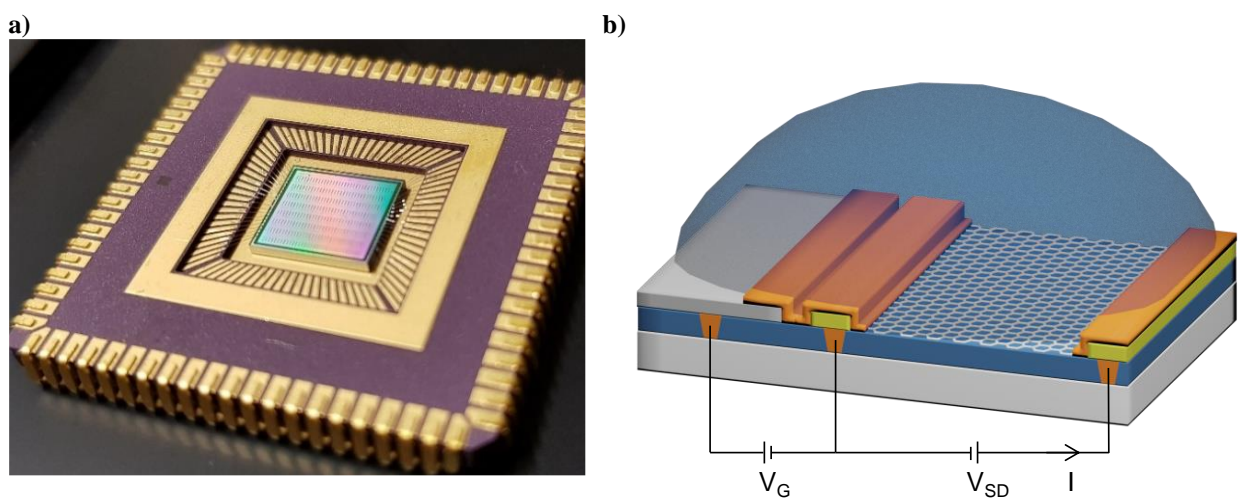

**Figure 3. a) A chip with CMOS integrated GFETs wire bonded on a chip carrier. b) Electrical measurement configuration of the GFETs. The on-chip platinum liquid gate electrode is used to control the liquid gate voltage ($V_G$). The current (I) in the graphene channel is measured between the source and drain electrodes with 0.1 V source-drain bias ($V_{SD}$) over the channel.**

The graphene detector surface was cleaned with isopropanol and DIW, before characterization of the devices in de-ionized water (DIW) and in a series of sodium chloride (NaCl) with different concentrations. For each concentration of NaCl, the sensor surface was further washed 5 times with a 100 μL droplet of the corresponding solution to ensure that the concentration on top of the chip was correct.

To evaluate the response of the sensor chip to the physiological solutions, the ionic strength of the sample solution was varied. The tests included measurements in DIW, 1 mM, 10 mM, and 100 mM NaCl solutions. The channel of the graphene FETs is charged, and an electrical double layer (EDL) is formed on the graphene-liquid interface to neutralize the charged surface. The thickness of the EDL can be estimated by using the Debye length which corresponds to the distance at which the electric potential has decreased in magnitude by 1/e. It can be calculated for the monovalent electrolytes at room temperature according to the following equation:

$$\lambda_D = 0.304/\sqrt{I},$$

where *I* is the ionic strength of the solution.

Based on the forum, the Debye lengths for the 1 mM, 10 mM, and 100 mM NaCl concentrations are 9.6 nm, 3.0 nm and 0.96 nm, respectively. The Debye length corresponds approximately to the gate oxide thickness of the GFETs. The dielectric constant of the DIW is 78.4[29,30] at 25 °C and approximately the same value applies for the low NaCl concentration solutions[31]. Thus, the change in the dielectric thickness of a GFET induced by the change in the NaCl concentration is expected to be seen as a shift in the Dirac peak position.

## 3. Results and discussion
**Graphene FET yield and uniformity**

In this work we focused on studying 512 GFETs with a resistive readout option from the five selected chips numbered from #1 to #5. The resistance values measured in ambient conditions without gating from chip #4 are shown in Figure 4a and histograms of the measured resistance values for 5 chips are in Figure 4b. The average resistance values and standard deviations (SD) are 790 Ω (SD=120 Ω), 830 Ω (SD=140 Ω), 810 Ω (SD = 120 Ω), 770 Ω (SD = 100 Ω) and 810 Ω (SD=160 Ω) for chips #1–5, respectively. This demonstrates good process stability over the whole wafer. A single non-contacted GFET was subtracted from the data for chips #3 and #5. This gives us an extremely high device yield of 99.9 % with 2558 devices working out of the 2560 measured devices.

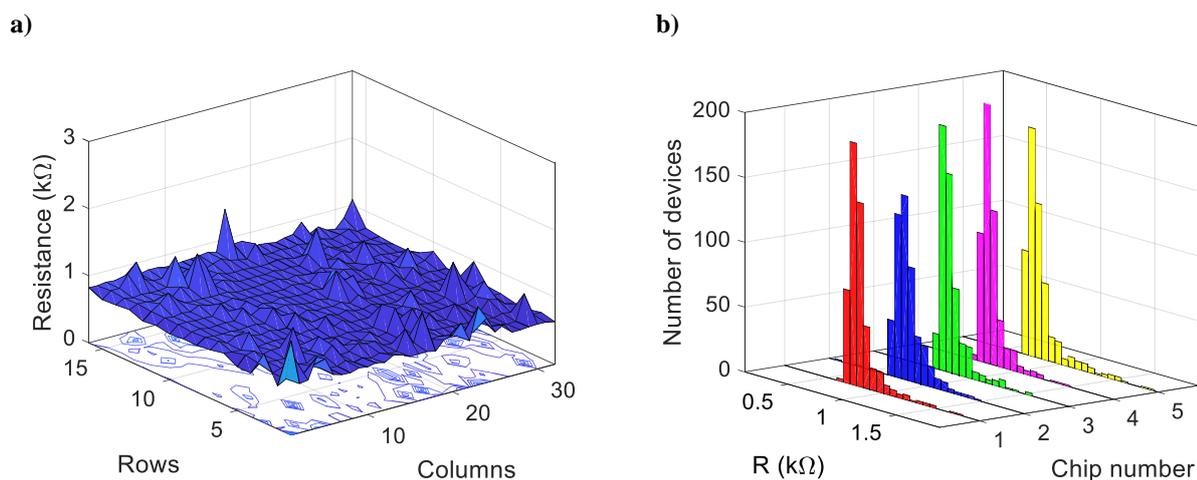

**Figure 4.** a) Resistance values of the 512 GFETs measured from the chip 4. The average resistance is 770 Ω (SD=100 Ω) b) Histograms of the resistances measured from the 5 chips. The average resistance values are 790 Ω (SD=120 Ω), 830 Ω (SD=140 Ω), 810 Ω (SD = 120 Ω), 770 Ω (SD = 100 Ω) and 810 Ω (SD=160 Ω) for chips #1–5, respectively. A single non-contacted GFET was subtracted from the data for chips #3 and #5. This gives us a very high device yield of 99.9 % with 2558 devices working out of the 2560 measured devices.

**Characterization in de-ionized water**

The performance of the GFETs was evaluated by electrical measurements. The GFETs were characterized in de-ionized water (DIW) to obtain resistance of the devices as a function of the gate voltage. The $V_G$ was applied using an on-chip Pt electrode to all the GFETs simultaneously. The results are shown in Figure 5a. The data has been normalized to the average Dirac peak voltage value of 1.45 V (SD = 0.01 V) measured in DIW. The average Dirac peak resistance is 6 kΩ (SD = 2 kΩ). The variation for the Dirac peak voltage is low but some of the devices show higher resistance levels, which is most likely caused either by a poor contact between the metal and graphene or possible defects in the graphene layer. Few devices also show a higher deviation in the Dirac peak voltage, which can possibly be caused by resist or passivation layer residues on top of the graphene channel. These residues are left from processing and cause additional doping and a change in the effective gate capacitance. Another possible reason for this is the use of DIW as a dielectric as it is possible that there are small air bubbles on the surface of some of the GFETs that interfere with the gating.

The transconductance of the GFETs is defined as the derivative of the channel current with respect to $V_G$. The transconductances as a function of the gate voltage are shown in Figure 5b. The transconductance maximum and minimum are 140 µS (SD = 30 µS) and -200 (SD = 30 µS), respectively. The corresponding voltage values for the maximum and minimum transconductance values are 200 mV (SD = 40 mV) and -160 mV (SD = 20 mV), respectively. The transconductance can be used to estimate the sensitivity of the devices in the biodetection. The attachment of the biomolecules typically induces gating to the graphene channel which causes the Dirac peak to shift. This shift can be measured as a change in the current versus time when measuring with a set gate voltage. The higher transconductance of the device then leads to a higher response in the measured current.

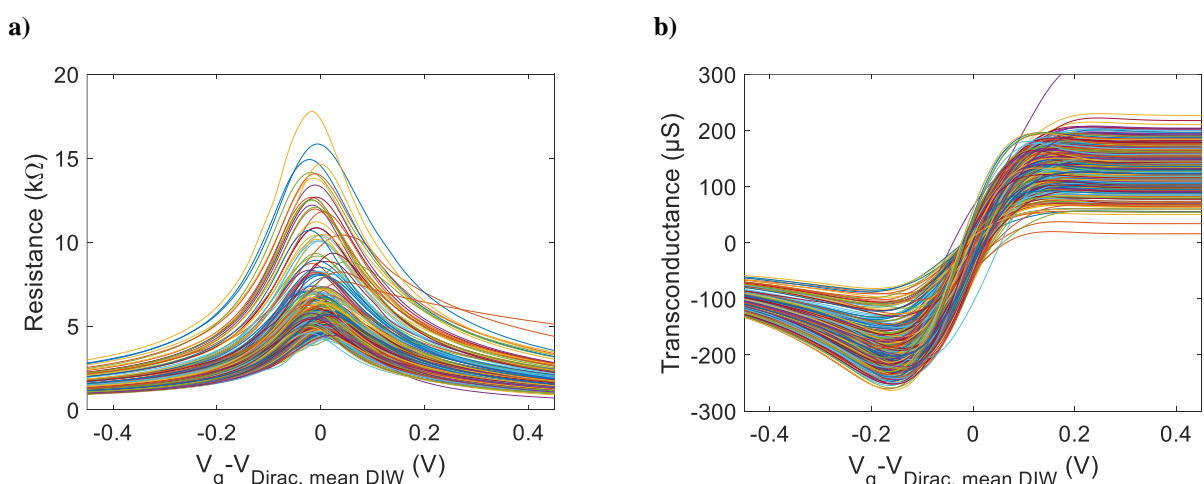

**Figure 5. a) Resistance values as a function of the $V_g$-$V_{Dirac,mean\ DIW}$ for the 512 GFETs. b) Transconductance values as a function of the $V_g$-$V_{Dirac,mean\ DIW}$ for the 512 GFETs.**

**Characterization in sodium chloride solutions**

To demonstrate the sensor sensitivity, the resistances of the GFETs as a function of $V_G$ were measured with a concentration series of NaCl in DIW. The tested NaCl concentrations in addition to DIW were 1 mM, 10 mM and 100 mM. The resistance values at Dirac peak and the corresponding Dirac peak voltage values for each concentration are shown in Figure 6a. The Dirac peak voltage that has been normalized to the average value in DIW changes from 0 mV (SD = 10 mV) in DIW to 45 mV (SD = 15 mV), 80 mV (SD = 15 mV) and 130 mV (SD = 20 mV) for 1 mM, 10 mM, and 100 mM NaCl concentrations, respectively. The data shows that the average resistance value at the Dirac peak stays stable at 6 kΩ (SD = 2 kΩ) when the concentration is changed, and the main signal is the shift of the Dirac peak voltage position. This indicates that the dominating sensing mechanism is electrostatic gating of the GFETs.[32] The sensitivity of each individual device has been estimated by using the measured Dirac voltage values and linear fitting shown in Figure 6b. The obtained average sensitivity of the GFETs for the NaCl concentrations in the 1 mM to 100 mM range is 42 mV/dec (SD = 4 mV/dec).

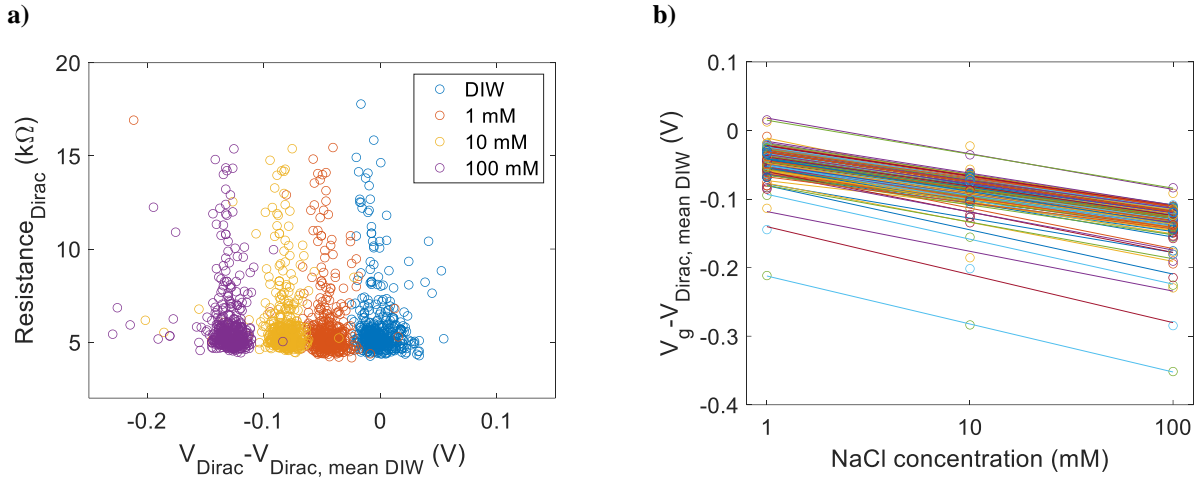

**Figure 6.** a) The resistance values at $V_{Dirac}$ as a function of $V_{Dirac}$-$V_{Dirac, mean\ DIW}$ for the 512 GFETs in deionized water, 1 mM, 10 mM and 100 mM sodium chloride solutions. b) The $V_{Dirac}$-$V_{Dirac, mean\ DIW}$ values as a function of NaCl concentration for the 512 GFETs used to extract the sensitivity of the devices in the 1 mM to 100 mM concentration range. The obtained average sensitivity of the GFETs for the NaCl solution in the 1 mM to 100 mM concentration range is 42 mV/dec (SD = 4 mV/dec).

## 4. Conclusions

We demonstrate, for the first-time, wafer-scale GFET CMOS integration for biosensing with device yield higher than 99% with a low device-to-device variability. We believe this approach will be crucial when commercializing GFET based biosensors, enabling multi-analyte sensing and the statistical analysis required for biologically quantitative on-chip bioassays. In the future, on-chip multi-analyte sensing could enable efficient screening of multiple viruses with sufficient statistics for reliable detection from a single small analyte sample. This technology enables the use of different sizes of GFET channels, with the possible number of GFETs ranging from thousands to millions of devices depending on the ASIC circuit design and technology. This flexibility in the number of and size of the devices allows for a wider set of applications where large amounts of biological information need to be evaluated, such as gene sequencing[33,34].

In addition to its application in quantitative bioassays, CMOS multiplexing can also be a valuable tool in the assay development phase. The most common issues in the development of FET based biosensing are related to the reliability and repeatability of individual devices and analysis. In many cases, the sensor chips are only used in a single measurement, hence individual faulty devices, voids in the functionalization and small air bubbles can lead to misinterpretation of results. The large-scale statistics provided by CMOS multiplexed sensor arrays on a single chip will improve the general reliability of the analysis and enables easy removal of defective devices from the data.

The other benefits that the monolithic integration of GFETs offers are reduction in environmental noise and simplified connections, when compared to hybrid off-chip sensors[35]. Furthermore, multiplexing sensor arrays offers advantages in sensor defect exclusion and wider dynamic ranges by allowing for varying sensor design for different sensitivities[33]. The demonstrated technology also opens possibilities for the wafer-scale fabrication of CMOS integrated GFET based gas sensor arrays[27] and infrared cameras[26] that have been demonstrated in the past. In this sense, the development of GFETs as a generic template that can be made specific by the choice of functionalization allows for the adoption of this technology in a wide range of different applications[36]. With CMOS integration, the readout can be tailored to fit specific needs, which leads to a truly versatile sensor platform.

## ASSOCIATED CONTENT

### AUTHOR INFORMATION

**Corresponding Author**

* E-mail: miika.soikkeli@vtt.fi

**Author Contributions**

The manuscript was written through contributions of all authors.

## ACKNOWLEDGMENT

The authors acknowledge the support by the European Union's Horizon 2020 research and innovation program under the grant agreements Graphene Flagship (881603) and 2D-EPL (952792) and the Academy of Finland (grant 342586) and Business Finland through co-innovation project DigiDNA (grant 1869/31/2016).